\documentclass[iop]{emulateapj}
\usepackage{apjfonts}
\usepackage{natbib,graphicx}
\usepackage{longtable}
\newcommand{\cp}{\citep}
\newcommand{\ct}{\citet}

\begin{document}

\title{The Role of Core Mass in Controlling Evaporation: The Kepler Radius Distribution and The Kepler-36 Density Dichotomy}

\author{Eric D. Lopez}
\author{Jonathan J. Fortney}
\affil{Department of Astronomy and Astrophysics, University of California, Santa Cruz, CA 95064} 

\begin{abstract}
We use models of coupled thermal evolution and photo-evaporative mass loss to understand the formation and evolution of the Kepler-36 system. We show that the large contrast in mean planetary density observed by \ct{Carter2012} can be explained as a natural consequence of photo-evaporation from planets that formed with similar initial compositions. However, rather than being due to differences in XUV irradiation between the planets, we find that this contrast is due to the difference in the masses of the planets' rock/iron cores and the impact that this has on mass loss evolution. We explore in detail how our coupled models depend on irradiation, mass, age, composition, and the efficiency of mass loss. Based on fits to large numbers of coupled evolution and mass loss runs, we provide analytic fits to understand threshold XUV fluxes for significant atmospheric loss, as a function of core mass and mass loss efficiency. Finally we discuss these results in the context of recent studies of the radius distribution of {\it Kepler} candidates. Using our parameter study, we make testable predictions for the frequency of sub-Neptune sized planets. We show that 1.8-4.0 $R_{\mathrm{\oplus}}$ planets should become significantly less common on orbits within 10 days and discuss the possibility of a narrow ``occurrence valley'' in the radius-flux distribution. Moreover, we describe how photo-evaporation provides a natural explanation for the recent observations of \ct{Ciardi2012} that inner planets are preferentially smaller within the systems.
\end{abstract}

\section{Introduction}

The Kepler-36 system \cp{Carter2012} is a fascinating case-study for theories of planet formation and evolution. It contains a closely spaced pair of super-Earth mass planets with periods of 13.8 and 16.2 days orbiting a slightly evolved $6.8\pm1.0$ Gyr old G1 sub-giant that is 2.9 times more luminous than the Sun. Although too faint for reliable radial velocity measurements, the system exhibits strong Transit Timing Variations (TTVs), which allowed the planet densities to be determined to better than 10\% precision \cp{Carter2012}. Surprisingly, despite their extremely similar orbits, the planets have densities that differ by almost an order of magnitude. The inner planet Kepler-36b has a mass of $4.5\pm0.3$ $M_{\mathrm{\oplus}}$ and a density of $7.2\pm0.7$ $\mathrm{g \, cm^{-3}}$, fully consistent with an Earth-like composition. Meanwhile the outer planet Kepler-36c has a mass of $8.1\pm^{0.6}_{0.5}$  $M_{\mathrm{\oplus}}$ but a density of only $0.87\pm0.05$ $\mathrm{g \, cm^{-3}}$, meaning that it must have a substantial H/He envelope \cp{Carter2012}.

This poses an interesting problem for theories of planet formation and evolution: how is it that two planets in the same system with very similar highly irradiated orbits arrived at such radically different densities today? One possibility is that the planets simply formed in very different environments. Models of core accretion show that it is much easier to accrete a substantial H/He envelope when the local disk temperature is lower \cp{Ikoma2012}. Perhaps Kepler-36b formed at or close to its current orbit while Kepler-36c formed substantially further out and migrated inwards \cp{Ida2010}. 

However, another possibility is that the planets did form in similar environments, but that subsequent evolution has caused them to diverge. In particular, photo-evaporation due to extreme ultra-violet (XUV) heating can remove large amounts of hydrogen/helium from highly irradiated planets through hydrodynamic mass loss. Models of XUV-driven mass loss were first developed to study water loss from early Venus \cp{Hunten1982, Kasting1983}, and hydrogen loss from the early Earth \cp{Sekiya1980, Watson1981}. These kinds of models were further developed to study mass loss from hot Jupiters \cp[e.g.,][]{Lammer2003, Yelle2004, Murray-Clay2009}, where there is clear evidence that atmospheric escape is an important physical process. This includes both trends in the population of hot Jupiters \cp{Lecavelier2007, Hubbard2007a, Hubbard2007b, Davis2009, Ehrenreich2011} and direct detections of atmospheric escape \cp{Vidal-Madjar2003, Vidal-Madjar2004, Linsky2010, Lecavelier2010, Lecavelier2012, Haswell2012, Ehrenreich2012}. More recently, mass loss models have been used to study the new populations of super-Earths and sub-Neptunes being found by transiting surveys like CoRoT and {\it Kepler} \cp{Jackson2012,Owen2012,Lopez2012,Wu2012,Owen2013}.

Recently, in \ct{Lopez2012} we showed that exoplanet compositions are subject to a photo-evaporation threshold.  Observationally there are no planets with low bulk density and high incident flux, implying that low-mass planets with substantial H/He envelopes do not exist in this area of parameter space.  Following \ct{Lecavelier2007} who studied hot Jupiters, \ct{Lopez2012} were able to show that this threshold can naturally be explained as a critical mass loss timescale.  Detailed models in \ct{Lopez2012}, which include coupled thermal and mass loss evolution, reproduced this threshold with standard mass loss efficiencies. We further showed that many of the super-Earths and Neptunes found by the {\it Kepler} mission including those in Kepler-11\cp{Lissauer2011a} and Kepler-36 lie along this threshold, indicating that these planets may have undergone substantial mass loss in the past. Here we show that photo-evaporation can be controlled by the mass of a planet's rock/iron core and that this provides a natural explanation for the divergent densities of Kepler-36 b\&c.  This process allows both planets to form with similar compositions and in similar environments before being sculpted by their different mass loss histories.

In addition to detailed studies of individual systems like Kepler-36, there is a growing body of literature examining the overall distribution of {\it Kepler} planet candidates. Detailed studies of planet occurrence rates by \ct{Petigura2013} and \ct{Fressin2013} have recently shown that is a  sharp drop off in the frequency of planets above $\gtrsim$2.8 $R_{\mathrm{\oplus}}$, at least within 50 and 85 days respectively. In contrast, planets with radii $\lesssim$2.8 $R_{\mathrm{\oplus}}$ seem to be equally common. Likewise, \ct{Howard2011a} and \ct{Youdin2011} showed that there is a drop off in the frequency of candidates at extremely short periods $\lesssim$10 days; moreover, this drop off is particularly acute for sub-Neptune sized planets. In multi-planet systems, \ct{Ciardi2012} showed that there is tendency for outer planets to be larger than inner planets in the same system, particularly when those planets are smaller than Neptune and on orbits $\lesssim$20 days. 

Models have shown that photo-evaporation and thermal evolution can significantly alter the H/He inventories of super-Earth and sub-Neptune sized planets \cp{Baraffe2006,Lopez2012,Owen2013}. Moreover, planet structure models have shown that changing the mass of a planet's H/He envelope has a large impact on its resulting radius \cp{Rogers2010b,Lopez2012}. As a result, models of thermal and mass loss evolution can shed light on many of the observed trends in the radius distribution of {\it Kepler} planet candidates.

\section{Our Model}
\label{modelsec}

For this work, we have used the coupled thermal evolution and mass loss model presented in \ct{Lopez2012}, where additional model details can be found. Similar models have been used to track the coupled evolution of rocky super-Earths \cp[e.g,][]{Jackson2010,Valencia2010,Nettelmann2011}, hot Neptunes \cp[e.g,][]{Baraffe2006}, and hot Jupiters \cp[e.g,][]{Baraffe2004,Baraffe2005,Hubbard2007a,Hubbard2007b}. Beginning shortly after the end of planet formation, we track planetary mass and radius as a function of age.  The use of coupled model is essential, because planetary radii are largest at young ages, when stellar XUV fluxes are highest.

At a given age, a model is defined by the mass of its heavy element core, the mass of its H/He envelope, the amount of incident radiation it receives, and the internal entropy of its H/He envelope.  Here we assume an isothermal rock/iron core with an Earth-like 2:1 rock/iron ratio, using the ANEOS olivine \cp{Thompson1990} and SESAME 2140 Fe \cp{Lyon1992} equations of state (EOS). For the H/He envelope we assume a fully adiabatic interior using the \ct{Saumon1995} EOS.

In order to quantitatively evaluate the cooling and contraction of the H/He envelope, we use a model atmosphere grid over a range of surface gravities and intrinsic fluxes. These grids relate the surface gravity and internal specific entropy to the intrinsic flux emitted for a given model. These radiative transfer models are computed at a Uranus and Neptune-like 50$\times$ solar metallicity atmosphere using the methods described in \ct{Fortney2007} and \ct{Nettelmann2011}. These atmosphere models are fully non-gray, i.e. wavelength dependent radiative transfer is performed rather than simply assuming a single infrared opacity. In addition, we include heating from radioactive decay in the rock/iron core and the delay in cooling due to the core's heat capacity.  In order to correctly determine a planet's mass loss history, it is vital to include these thermal evolution effects, since these will strongly affect a planet's radius over time. Radius, in turn, has a large impact on the mass loss rate as seen in Equation \ref{masslosseq}.


Close-in planets like those in Kepler-36 are highly irradiated by extreme ultraviolet (EUV) and X-ray photons. These photons photo-ionize atomic hydrogen high in a planet's atmosphere, which in turn produces significant heating \cp{Hunten1982}. If this heating is large enough, it can generate a hydrodynamic wind that is capable of removing significant mass. We couple this XUV-driven mass loss to our thermal evolution models using the energy-limited approximation \cp{Watson1981}. This allows a relatively simple analytic description of mass loss rates. 

\begin{equation}\label{masslosseq}
\dot{M}_{\mathrm{e-lim}} \approx \frac{\epsilon \pi F_{\mathrm{XUV}} R_{\mathrm{XUV}}^3}{G M_{\mathrm{p}} K_{\mathrm{tide}}}
\end{equation}
\begin{equation}
K_{\mathrm{tide}} = (1 - \frac{3}{2 \xi} + \frac{1}{2 \xi^3})
\end{equation}
\begin{equation}
\xi = \frac{R_{\mathrm{Hill}}}{R_{\mathrm{XUV}}}
\end{equation}

Equation (\ref{masslosseq}) describes our estimate of the mass loss rate based on the formulation from \ct{Erkaev2007}. $F_{\mathrm{XUV}}$ is the time-dependent total flux between $1-1200$ \AA, which is given by \ct{Ribas2005} as a function of age for Sun-like stars. $R_{\mathrm{XUV}}$ is the planetary radius at which the atmosphere becomes optically thick to XUV photons, which occurs at pressures around a nanobar \cp{Murray-Clay2009}. $K_{\mathrm{tide}}$ is a correction factor that accounts for the fact that mass only needs to reach the Hill radius to escape \cp{Lecavelier2004, Erkaev2007}. Finally, $\epsilon$ is an efficiency factor that parametrizes the fraction of the incident XUV flux that is converted into usable work. For this work we use $\epsilon=0.1$ based on the observed photo-evaporation threshold described in \ct{Lopez2012}. This value is similar to the efficiencies found by \ct{Owen2012}. Using more sophisticated photo-evaporation models for hot-Neptunes they found mass loss efficiencies that varied from 0.05 to 0.2. We make an additional conservative modeling choice by starting mass loss at an age of 10 Myr, since stellar XUV fluxes and planetary radii post-formation are not well understood at even earlier times.

\section{Kepler-36: Explained by Mass Loss?}
\label{k36sec}

\begin{figure*}
  \begin{center}
    \includegraphics[width=6.7in,height=4.47in]{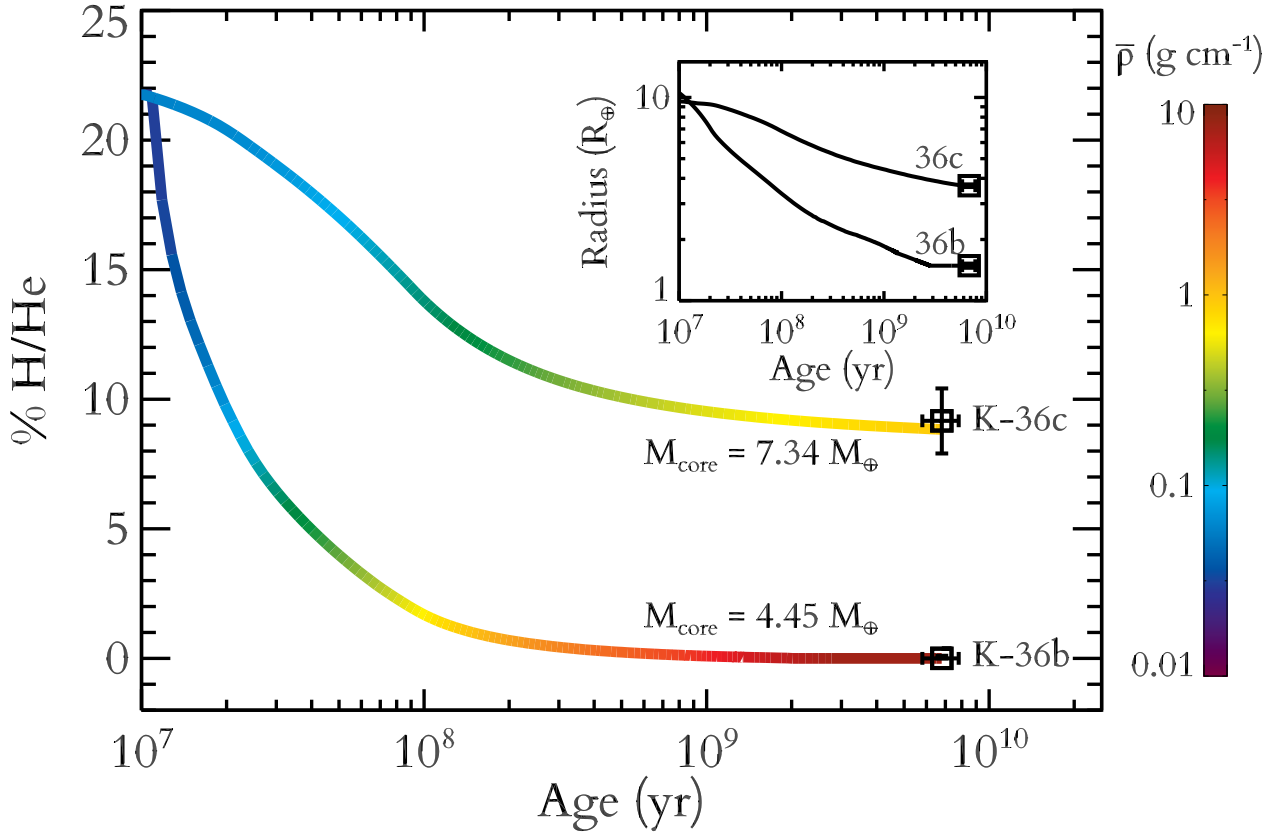}
  \end{center}
  \caption{Possible mass loss histories for Kepler-36b \& c. The curves plot the mass fraction in the H/He envelope vs. time, while the colors indicate each planet's density at a given age. Likewise the inset shows the radius evolution of each planet. The black points on the right hand side indicate the current compositions of Kepler-36b \& c, while the text lists the rock/iron mass predicted for each planet. Currently, Kepler-36c requires $\sim 8 \%$ of its mass in H/He, while Kepler-36b is consistent with an Earth-like composition. Nonetheless it is possible both planets formed with $\sim 22 \%$ H/He, but subsequent mass loss has distinguished them due to differences in their core masses. This provides a natural explanation for the large density contrast seen today in this system.  Such evolutionary histories are a generic outcome of our evolution calculations, with a wide range of initial H/He masses and mass loss efficiencies.\label{k36fig}}
\end{figure*}

Before we determine the possible mass loss histories of Kepler-36b \& c, we must first examine their present day compositions. For Kepler-36b, this is relatively straightforward since its density is consistent with a rocky compositions without a H/He envelope. In this case we find that it should be $25.9\pm^{9.1}_{4.3}$ \% iron, consistent with an Earth-like rock/iron ratio. This is derived by matching the observed mass and radius with our structure models in the absence of any H/He or water envelope. Likewise, the error bars include the observed 1$\sigma$ uncertainties in mass and radius. In contrast, Kepler-36c has much lower density and requires a substantial H/He envelope to explain its radius. To estimate its present day composition, we ran our thermal evolution models in the absence of any mass loss. To explain its current radius Kepler-36c must be $8.6\pm1.3\%$ H/He, assuming an Earth-like core. To calculate the error bars we varied the planetary albedo from 0-0.8 and the heat capacity of the rocky core from 0.5-1.0 $\mathrm{J \, K^{-1} \, g^{-1}}$. We also included the observed uncertainties in mass, radius, current age, and incident flux. In addition, for Kepler-36b we can set an upper limit on the presence of any H/He envelope.  It must be $<0.1\%$ H/He, assuming a maximally iron rich core \cp{Marcus2010}.

By knowing the present day compositions, we can determine the mass of each planet's core. For Kepler-36b it is simply the observed mass from TTVs, $4.5\pm0.3$ $M_{\mathrm{\oplus}}$, while for Kepler-36c it is $7.4\pm^{0.5}_{0.4}$ $M_{\mathrm{\oplus}}$. Using these core masses, we ran fully coupled models including both mass loss and thermal evolution, in order to determine the initial composition each planet had after formation. To ensure consistency, we check that these models with mass loss are still able to reproduce present day radius. 

Figure \ref{k36fig} shows the results of our coupled mass and thermal evolution models for both planets in the Kepler-36 system.   The model assumes that both planets formed at their current orbits with the same initial H/He mass fraction. Each curve plots the fraction of each planet's mass in the H/He envelope vs. age, while the colors indicate the planet's density at that age. The black points at the right indicate the current age and composition of the Kepler-36 planets. We also show the radius evolution of both planets in an inset. Such large radii at young ages are a generic outcome of evolution models, since the H/He envelopes are warm and are not degenerate. See \ct{Mordasini2012c} for a wider exploration of planetary radii for low-mass planets with H/He envelopes as a function of age.

At 10 Myr, when we start photo-evaporation both planets are $\sim 10$ $R_{\mathrm{\oplus}}$. Kepler-36b rapidly contracts as it loses mass and is down to $\sim 3$ $R_{\mathrm{\oplus}}$ by 100 Myr. Kepler-36c is also vulnerable to mass loss. To have retained its current 8\% of H/He Kepler-36c would need to have been formed with 22\% H/He at 10 Myr and a mass of 9.4 $M_{\mathrm{\oplus}}$. Since Kepler-36b is consistent with no H/He envelope today, we can only set upper limits on its initial composition. However, if we assume that it formed with the same initial 22\% H/He as 36c, then it would have lost its entire H/He envelope by the time it was 2 Gyr old. Thus the large discrepancy in the present day densities of the two planets, can naturally be explained by the fact that Kepler-36b is significantly more vulnerable to mass loss than Kepler-36c.

\begin{figure*}[t] 
  \begin{center}
    \includegraphics[width=7.in,height=4.0in]{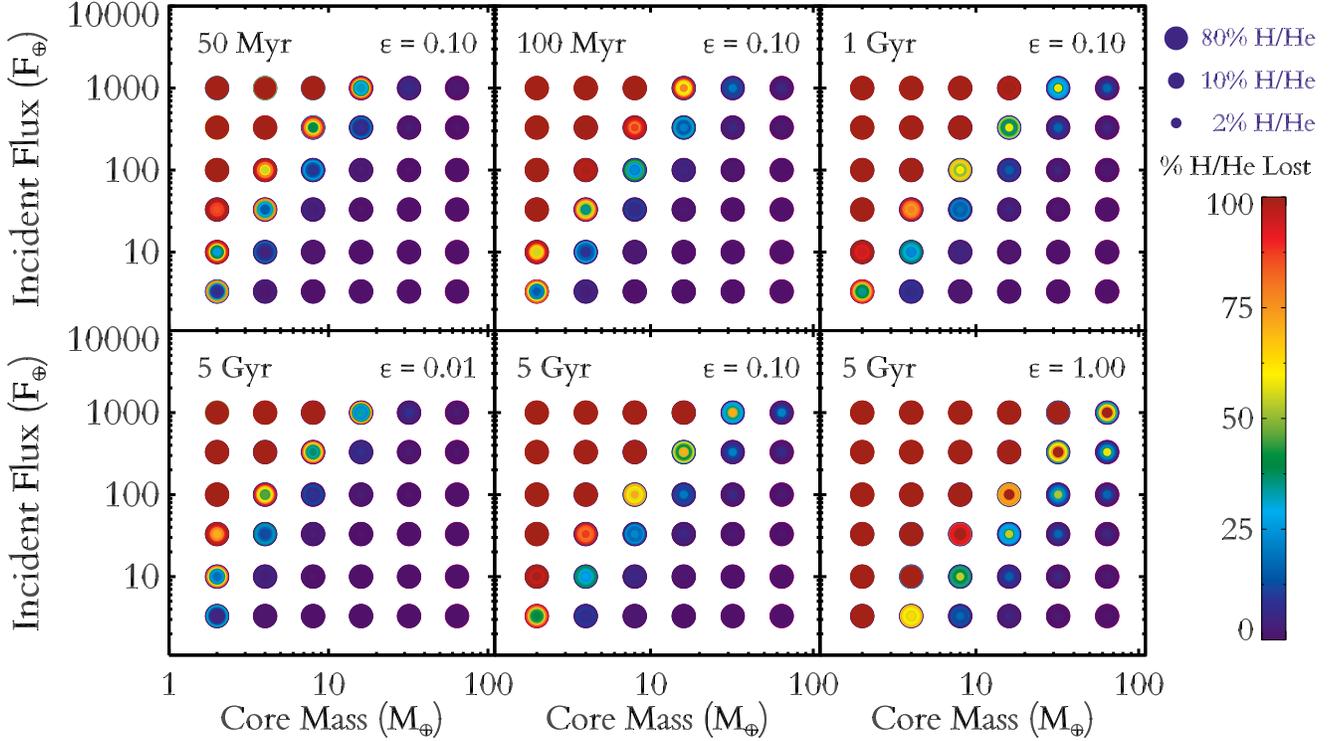}
  \end{center}
  \caption{This summarizes the results of $\sim$ 6000 model runs designed to explore how our coupled thermal and mass loss evolution models depend on basic planet properties. In each panel we have plotted models on a grid of incident bolometric flux, relative to the flux that the Earth receives from the Sun, vs. planetary core mass in Earth masses. At each point in the core mass-flux grid, we have over-plotted multiple models with different initial compositions. The size of each point indicates the initial H/He fraction a planet began with. Meanwhile, colors indicate how much of this initial H/He envelope is lost over time. Thus indigo and blue points are models where mass loss is unimportant while dark red points are models where the entire H/He envelope has been stripped off. Clearly, the mass loss history depends strongly on a planet's core mass as well as the incident flux it receives. In each case, there is a clear threshold region where mass loss is important but, at least some models are able retain significant H/He envelopes. This threshold region can be described roughly as a power-law with the threshold flux, $F_{\mathrm{th}}\propto M_{\mathrm{core}}^{2.4}$. See Figure \ref{fit1fig}a and equation (\ref{thresholdeq}). The different panels show how the location of this threshold varies with age and mass loss efficiency. The top panels show the threshold at 50 Myr, 100 Myr, and 1 Gyr for our standard mass-loss efficiency of 10\%. Most of the mass is lost in the first 100 Myr while almost no mass is lost after 1 Gyr. Likewise, the bottom panels show the results at 5 Gyr for efficiencies of 1, 10, and 100\%. Models with higher mass loss efficiencies are naturally more vulnerable to mass loss.\label{studyfig}}
\end{figure*}

However, this leads to the question, why is Kepler-36b so much more vulnerable than its neighbor? Although it does receive $\approx 24 \%$ more incident radiation than 36c, this alone produces far too small an effect to explain the necessary difference in mass loss histories \cp{Carter2012}. Instead, it is caused almost entirely by the difference in the masses of the two planets. As shown in \ct{Lopez2012}, planetary radius at fixed composition is relatively flat as a function of mass for $\sim 1-20$ $M_{\mathrm{\oplus}}$ planets with significant H/He envelopes. As a result, density for these planets goes roughly like the mass. Since the mass loss rate is proportional to the inverse of the of the average density $1/\bar{\rho}$, the mass loss timescale goes like $M_{\mathrm{p}} \times \bar{\rho}$ or $M_{\mathrm{p}}^2$. 

However, rather than considering a planet's total mass today, it is useful to consider simply its heavy element mass today. Unlike the current total mass and bulk density, the heavy element mass should remain essentially constant as a planet contracts and loses its H/He envelope, providing a useful tool for estimating a planet's mass loss rate throughout its history. For simplicity, we assume here that the heavy elements are locked in a silicate/iron core, although it is also possible that some of these metals could be mixed into the H/He. Of course at a given composition, the mass of this rocky core $M_{\mathrm{core}}$ is proportional to the total mass $M_{\mathrm{p}}$. Thus the mass loss timescale $t_{\mathrm{ml}}$ goes roughly like $M_{\mathrm{core}}^2$. We calculate that Kepler-36c's rocky core should be $65\%$ more massive than that of Kepler-36b. As a result, we predict that the mass loss timescale for Kepler-36b was $\sim 3$ times shorter than for Kepler-36c.

In Figure \ref{k36fig}, we have presented one possible scenario for the evolution of Kepler-36. These results are insensitive to any reasonable variation in the mass loss efficiency or the initial envelope fractions. For mass loss efficiencies significantly less than 0.1, it is not possible to construct a scenario where both planets b and c started out with identical envelope fractions. Nonetheless, even for lower efficiencies it is still quite possible that both planets started off with substantial amounts of H/He but that Kepler-36b lost its gaseous envelope to mass loss.

\section{The Role of Core Mass}
\subsection{A Parameter Study}
\label{studysec}

The Kepler-36 system clearly demonstrates that the mass loss evolution of planet depends on more than just the incident XUV flux that a planet receives. In \ct{Lopez2012} we showed that there is a threshold in the observed population of transiting planets with measured densities and H/He envelopes. This threshold is well described by a critical mass loss timescale:

\begin{equation}
t_{\mathrm{loss}} =  \frac{G M_{\mathrm{p}}^2}{\pi \epsilon R_{\mathrm{p}}^3 F_{\mathrm{XUV,E100}}} \frac{F_{\mathrm{\oplus}}}{F_{\mathrm{p}}} \approx 12 Gyr 
\end{equation}

Here $F_{\mathrm{p}}$ is the incident bolometric flux that a planet receives from its parent star,  $F_{\mathrm{\oplus}}$ is the current bolometric flux that the Earth receives from the Sun, and $F_{\mathrm{XUV,E100}} = 504$ $\mathrm{erg \, s^{-1} \, cm^{-2}}$ is the XUV flux at the Earth when it was 100 Myr old. In \ct{Lopez2012} we then performed a simple parameter study to show that this critical mass loss timescale was well reproduced by our coupled thermal and mass loss evolution models. 

In order to better understand in detail how mass loss evolution depends on the mass of a planet's rocky core as well as the incident flux and mass loss efficiency, we greatly expanded the parameter study from \ct{Lopez2012}. Beginning at 10 Myr, we ran over $6000$ evolution models varying the incident flux, core mass, composition, and mass loss efficiency. We then recorded the radius, mass, and composition at different ages. The grid of initial conditions was evenly spaced logarithmically, with incident flux varying from $1-1000$ $F_{\mathrm{\oplus}}$, the rocky core mass from 1-64 $M_{\mathrm{\oplus}}$, the initial H/He mass fraction from 0.1-80\% H/He, and the mass loss efficiency from 0.01-1.0. Figure \ref{studyfig} summarizes the results. In each panel we have plotted the incident flux vs. the core mass. Each circle corresponds to an individual planet that is color-coded by the fraction of its initial H/He envelope that is lost by the age indicated on the panel. At each point we overplotted multiple models with different initial compositions, in each case the size of the circle corresponds to the initial mass fraction in the H/He envelope. Finally the different panels compare results at different times and for different mass loss efficiencies. The top three panels show the results using our standard mass-loss efficiency at 50 Myr, 100 Myr, and 1 Gyr. Meanwhile the bottom three panels show the results at 5 Gyr, a typical age for \emph{Kepler} systems, for mass loss efficiencies of 0.01, 0.1, and 1.0.

Clearly, the mass loss history of a planet depends strongly on both the incident flux and the mass of the rock/iron core. The indigo models in the lower right of each panel have lost a negligible fraction of their initial H/He envelope. These models have either relatively massive cores and/or receive little incident flux and so mass loss is unimportant to the evolution of planets in this part of parameter space. On the other hand, the dark red models in the upper left with low mass cores and high incident flux have completely lost their entire H/He envelopes. This region of parameter space should be filled with highly irradiated rocky planets like CoRoT-7b and Kepler-10b \cp{Leger2009,Queloz2009,Batalha2011}. 

In between there is a transition region where mass loss is important but at least some of the models are able to retain a H/He envelope. This transition region is relatively narrow, spanning less than an order of magnitude in incident flux for a given core mass. For models in this transition region the relation between the size of the initial H/He envelope and the fraction of the envelope lost is extremely complicated and not always monotonic. There is trade off between the fact that planets with small initial envelopes have relatively little mass in those envelopes to lose, and the fact that planets with large initial envelopes have larger initial radii and therefore experience much higher mass loss rates. For models at early times and/or low mass loss efficiency, the fraction of the envelope lost increases with the initial envelope fraction. On the other hand, at late times and/or high mass loss efficiencies it is the planets with small initial envelopes that are most vulnerable. As a result, much of the scatter in characterizing this transition region is determined by variations in the initial H/He envelope fraction.

\begin{figure}[h] 
  \begin{center}
    \includegraphics[width=3.5in,height=7.4in]{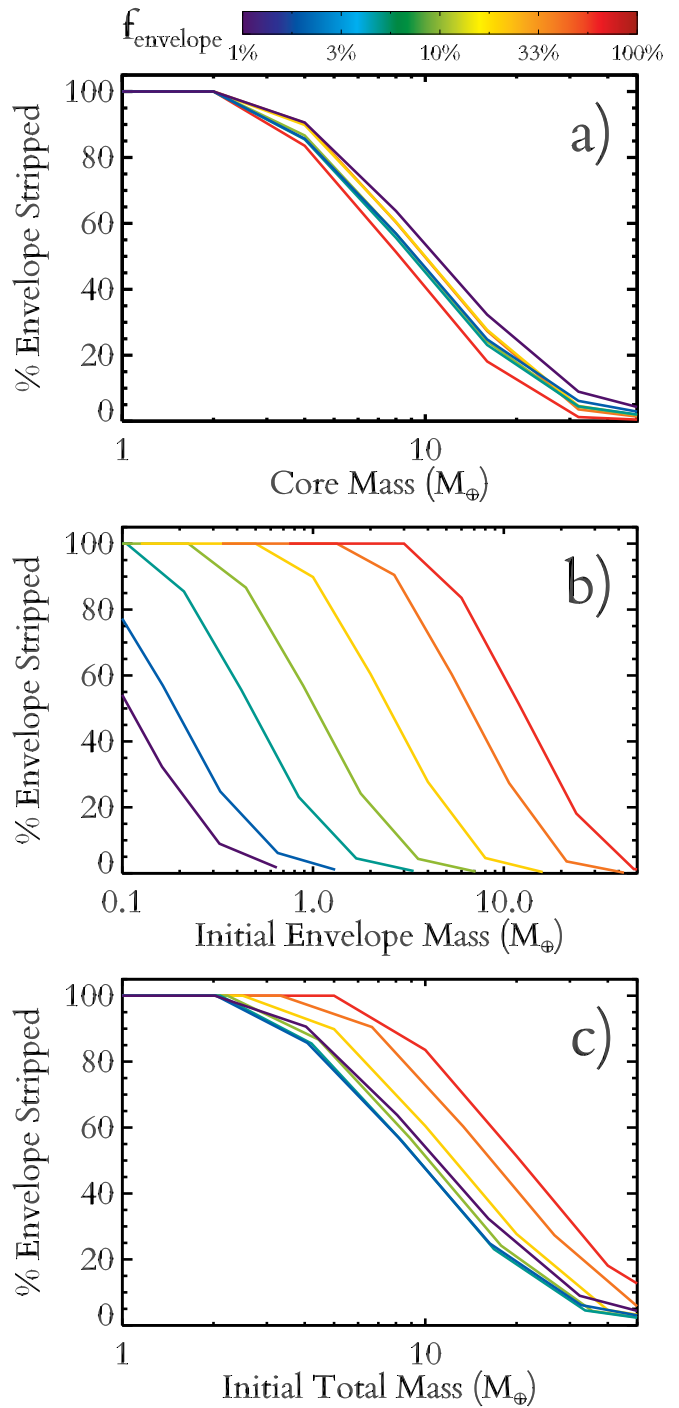}
  \end{center}
  \caption{The fraction of the initial H/He envelope which is lost to subsequent photo-evaporation ($f_{\mathrm{lost}}$) vs. three different mass parameters, according to our models. Panel a) plots $f_{\mathrm{lost}}$ against the mass of the rock/iron core while panel b) plots it against the mass of the initial H/He envelope and panel c) plots against the total initial mass. In each case, the different curves show the results for models with different initial H/He fractions $f_{\mathrm{envelope}}$, varying from 1-60\% H/He. All of these models receive 100 $F_{\mathrm{\oplus}}$, assume $\epsilon = 0.1$, start at 10 Myr, and end at 5 Gyr. Compared to envelope or total mass, core mass shows the least scatter between models with different composition. \label{provefig}}
\end{figure}

In addition, the bottom three panels of Figure \ref{studyfig} make it clear that the location of this mass loss transition region depends on the mass loss efficiency. This makes sense; it is of course easier to remove more mass if the photo-evaporation is more efficient. Below, we will show that this behavior can also be well approximated as a power-law. In contrast, the behavior with age is more like an exponential decay. Most of the mass is lost in the first 100 Myr, while a comparison of the top right and bottom middle panels shows that relatively little mass is lost after 1 Gyr.

Likewise, in each panel the threshold for significant mass loss can be described as a power-law. If we define the criterion for significant mass loss to be the evaporation of half the initial H/He envelope by several Gyr, then at a given age and mass loss efficiency this defines a power-law relation between the rock/iron core mass and the threshold incident flux needed for substantial mass loss. The slope of this power-law is roughly $F_{\mathrm{th}} \propto M_{\mathrm{core}}^{2.4}$, as discussed in section \ref{scalesec}.

\subsection{Why Core Mass?}

The virtue of studying mass loss trends vs. core mass is that it allows us to define a narrow band of parameter space in Figure \ref{studyfig} where the precise details of mass loss are important with relatively little scatter due to variations in the initial H/He envelope mass. The currently observable planet properties, such as present day planet mass, radius, and density, are themselves highly dependent on the mass loss history and therefore on other unknown variables like the mass loss efficiency and initial composition. This makes it difficult to separate the effects of variations in current planet mass from variations in other parameters like incident flux or mass loss efficiency. On the other hand core mass represents an initial condition which is constant throughout a planet's mass loss history.

Rather than core mass we could alternatively chose to study trends against other theoretical parameters like the initial envelope mass or initial total mass after formation. Like core mass, these parameters are independent of any subsequent mass loss evolution; however, in our models these parameters also less adept at predicting that mass loss evolution. In Figure \ref{provefig} we show how $f_{\mathrm{lost}}$, the fraction of a planet's initial H/He envelope that is lost, depends on the core mass, the initial envelope mass, and the total mass. Each of the different curves is for a different initial composition $f_{\mathrm{envelope}}$ varying from 1-60\% H/He. All the models are computed at 100 $F_{\mathrm{\oplus}}$, assume $\epsilon=0.1$, start at 10 Myr, and end at 5 Gyr.

Ideally we want a parameter that minimizes the scatter between models with different initial compositions. Initial composition is not directly observable, and determining a planet's mass loss history is a much more model-dependent exercise than determining its current composition. For observed planets like those in Kepler-36, it is possible to tightly constrain the mass in a H/He envelope, even without the coupled thermal evolution models used here \cp{Rogers2010b}. Moreover, as we show in Figures \ref{fit1fig}-\ref{fit4fig}, trends with composition are complicated and cannot be described by a simple power-law. Thus parameters that minimize the scatter between models with different initial compositions are much better predictors of mass loss evolution. Comparing panels a) and b) in Figure \ref{provefig}, it is clear that plotting $f_{\mathrm{lost}}$ vs. core mass produces vastly less scatter than plotting $f_{\mathrm{lost}}$ vs. initial envelope mass. In panel c) we see that initial total mass is a reasonable predictor of $f_{\mathrm{lost}}$, but with about twice as much scatter as in panel a). Unique among a planet's properties, $M_{\mathrm{core}}$ is unchanged throughout a planet's evolution, is relatively model independent, and is a strong predictor of mass loss evolution.

\subsection{Scaling Relations for Coupled Mass Loss Evolution}
\label{scalesec}

\begin{figure}
  \begin{center}
    \includegraphics[width=3.0in,height=7.71in]{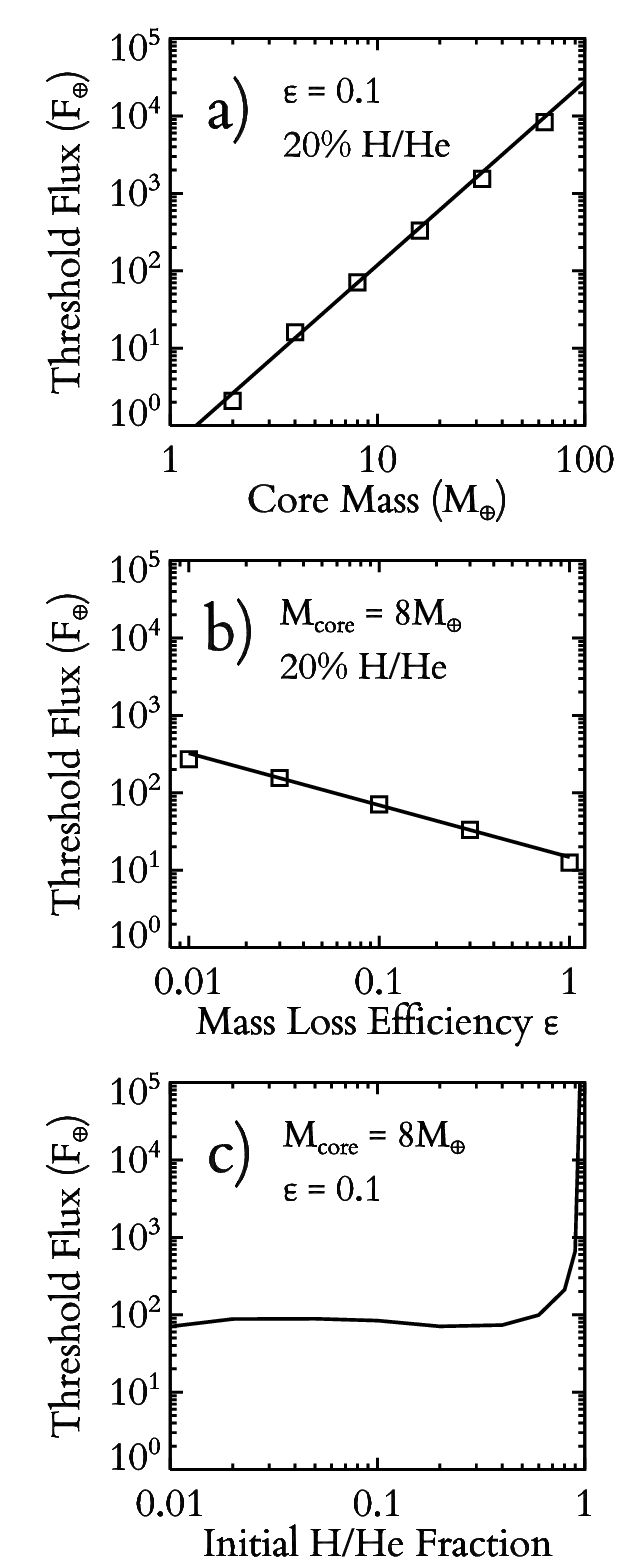}
  \end{center}
  \caption{Three panels showing how the threshold flux varies with core mass, mass loss efficiency and the initial composition. In panel a) we show how the threshold flux, the incident flux needed to remove half a planet's initial H/He envelope, varies with the core mass along with the best fit power-law. In panel b) we do the same for the mass loss efficiency. In panel c) we show how the threshold flux depends on the initial H/He envelope fraction. There is little dependence on composition until the planet starts out > 60\% H/He. \label{fit1fig}}
\end{figure}

In order to quantify the dependency of mass loss on initial conditions, we examine how the location of the mass loss transition region varies with each variable independently. This allows us to understand the qualitative behavior of our complete model in terms of a few simple scaling relations. Such relations can be used for quick and relatively accurate estimates of the importance of mass loss for detected planets.

We start by defining $F_{\mathrm{th}}$ as the threshold flux for a model with a given core mass, mass loss efficiency, age, and initial composition to lose half of its initial H/He envelope. Holding all other variables fixed we examine how $F_{\mathrm{th}}$ varies with each parameter and try fitting a power-law relation. We then vary the other parameters across our entire parameter study and examine the scatter in these power-law fits. The dependence of $F_{\mathrm{th}}$ on $M_{\mathrm{core}}$, $\epsilon$, and $f_{\mathrm{lost}}$ (the fraction of the initial H/He envelope that is lost) are all well fit by power-laws. As we previously described, the age dependence can be described as an exponential decay. For a 1 $M_{\mathrm{\oplus}}$ core and $\epsilon=0.1$, the best fit exponential decay is: 

\begin{equation}\label{decayeq}
F_{\mathrm{th}}\sim \exp{(-(t - 140 \,\, \mathrm{Myr})/80 \,\, \mathrm{Myr}) } F_{\mathrm{\oplus}} + 3.4 F_{\mathrm{\oplus}} 
\end{equation}

\noindent Meanwhile for systems older than 1 Gyr, age dependence is unimportant, allowing us to study $F_{\mathrm{th}}$ independently of age.

\begin{figure}
  \begin{center}
    \includegraphics[width=3.0in,height=2.57in]{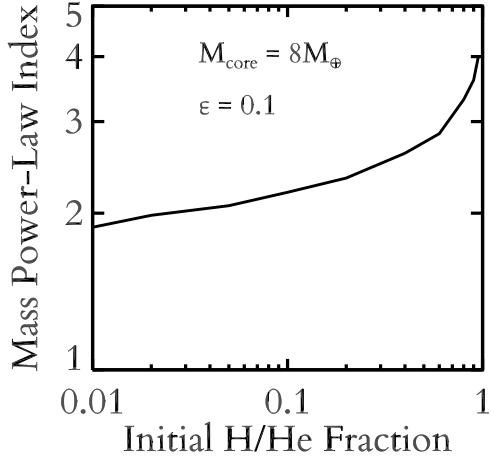}
  \end{center}
  \caption{Index of the core mass power-law from Figure \ref{fit1fig} and equation (\ref{thresholdeq}) vs. initial H/He envelope fraction. The dependence on core mass becomes steeper as the initial H/He fraction increases.\label{fit2fig}}
\end{figure}

\begin{figure}
  \begin{center}
    \includegraphics[width=3.0in,height=2.57in]{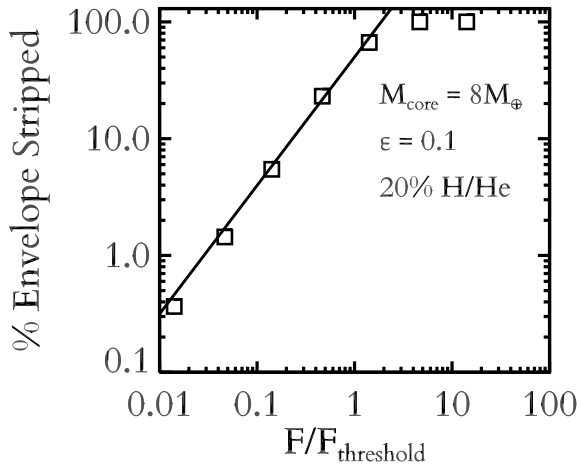}
  \end{center}
  \caption{$f_{\mathrm{lost}}$ the fraction of the initial H/He that is stripped after 5 Gyr vs. incident flux compared to $F_{\mathrm{th}}$ the threshold flux defined in equation (\ref{thresholdeq}). Below $F_{\mathrm{th}}$, $f_{\mathrm{lost}}$ increases roughly linearly with incident flux. Above $\sim$ 2$\times$ $F_{\mathrm{th}}$, the envelop is completely stripped.\label{fit3fig}}
\end{figure}

\begin{figure}
  \begin{center}
    \includegraphics[width=3.0in,height=2.57in]{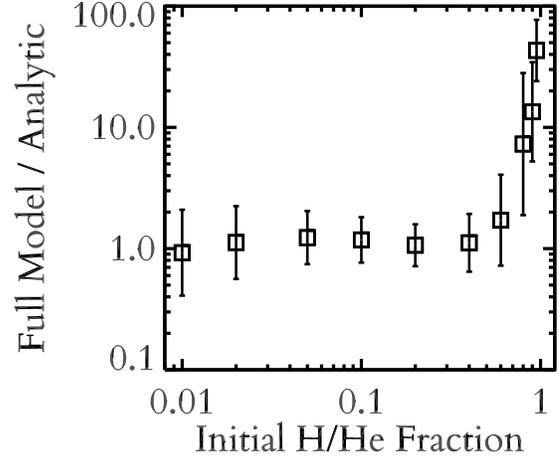}
  \end{center}
  \caption{$f_{\mathrm{lost}}$ according to the results of our full model from Figure \ref{studyfig} is divided by $f_{\mathrm{lost}}$ according to the simple analytic description in equation (\ref{thresholdeq}) and plotted against initial H/He fraction. For initial compositions < 50\% H/He, the two generally agree within a factor of 2 or better. \label{fit4fig}}
\end{figure}

Figure \ref{fit1fig} shows in detail how the threshold flux and the amount of H/He lost depend on each variable in our parameter study according to our thermal and mass loss evolution model. For clarity we have picked representative values for the core mass, mass loss efficiency, and the initial composition, but the results shown are generalizable across the entire parameter space. The default values correspond roughly to those for Kepler-36c: an 8 $M_{\mathrm{\oplus}}$ core, $\epsilon = 0.1$, and an initial composition of 20\% H/He. Likewise, we choose models where 50\% of a planet's initial envelope has been lost. We then vary $M_{\mathrm{core}}$, $\epsilon$, and composition one at a time. For clarity, in each panel we list the variables that are being held constant.

In Figure \ref{fit1fig}a we examine how $F_{\mathrm{th}}$ depends on $M_{\mathrm{core}}$. This is well described by the over-plotted power-law with $F_{\mathrm{th}} \propto M_{\mathrm{core}}^{2.4}$. This power-law is closely related to the critical photo-evaporation timescale that we described in \ct{Lopez2012}, where the mass loss timescale $t_{\mathrm{ml}}$ goes like $M_{\mathrm{p}}\rho/F_{\mathrm{p}}$. This corresponds to the transition region in Figure \ref{studyfig}. Because super-Earths and sub-Neptunes contain most of their mass in a rock/iron core, $\rho$, $M_{\mathrm{p}}$, and $M_{\mathrm{core}}$ all correlate strongly with each other, which implies that $F_{\mathrm{th}}\propto M_{\mathrm{p}}\bar{\rho} \propto M_{\mathrm{core}}^2$. Our fitted power-law, $F_{\mathrm{th}} \propto M_{\mathrm{core}}^{2.4}$, is slightly steeper than the simple $M_{\mathrm{p}}^2$ dependence we would expect analytically. The difference is due to the slight dependence of radius on core mass. 

\begin{figure*}
  \begin{center}
    \includegraphics[width=7.in,height=4.67in]{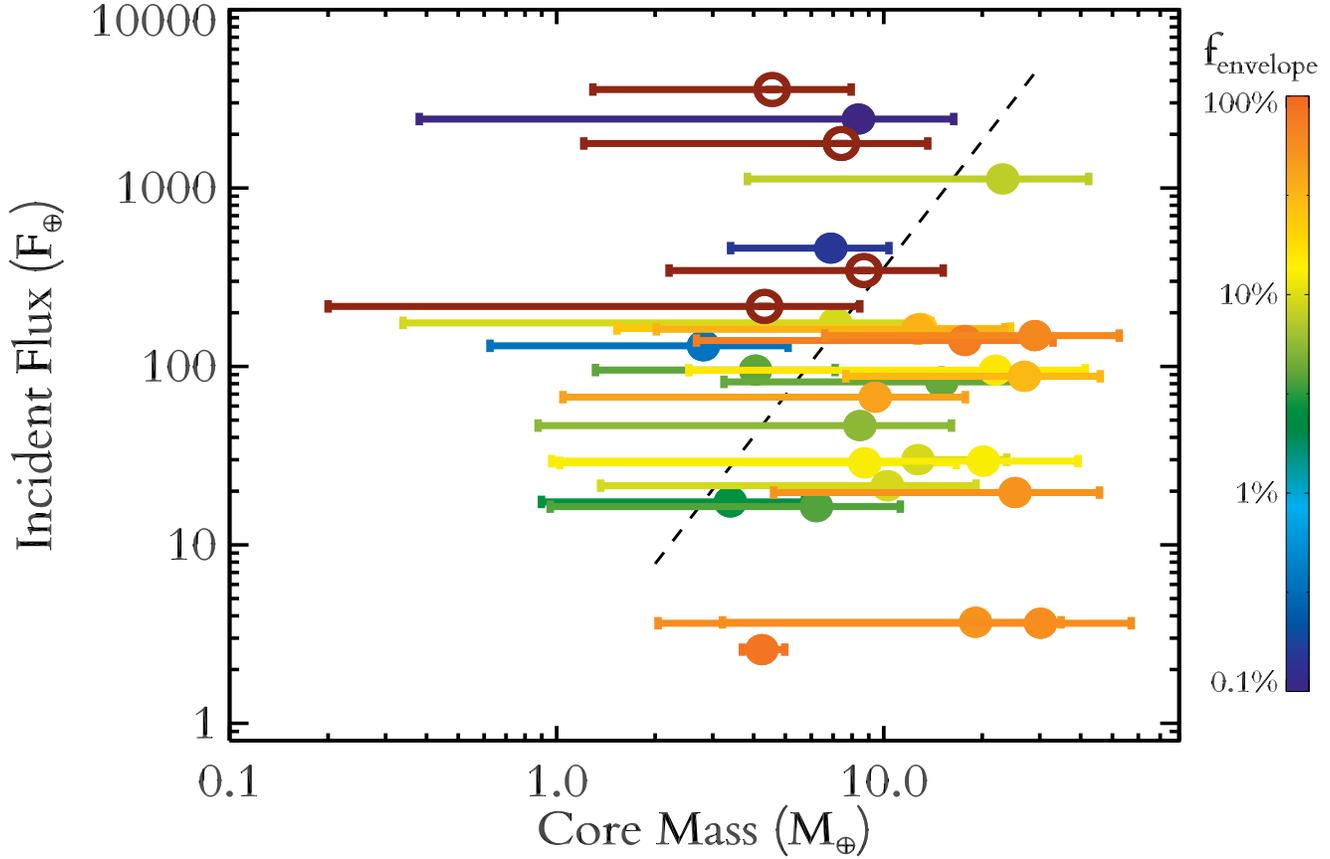}
  \end{center}
  \caption{Incident flux and core mass for 29 observed transiting planets with well defined masses less than 100 $M_{\mathrm{\oplus}}$. Planets are color-coded by their current H/He envelope fraction. Red-brown open circles are consistent with being H/He free. The dashed line shows the $F_{\mathrm{th}}-M_{\mathrm{\oplus}}$ relation from equations (\ref{thresholdeq}) and (\ref{losteq}), scaled up slightly to account for complete stripping of H/He. Of the six planets that lie 1 $\sigma$ to left of this relation, three are consistent with being rocky and three with having only water/steam envelopes.\label{corefig}}
\end{figure*}

Likewise in Figure \ref{fit1fig}b we examine how $F_{\mathrm{th}}$ depends on $\epsilon$. This is also well fit by a power-law, however it is intriguing that the dependence on $\epsilon$ is not quite inversely linear as it would be in a simple mass loss timescale. This is one of the effects of coupling mass loss and thermal evolution. All other things being equal, as the mass loss efficiency increases, the mass loss timescale becomes shorter while the thermal cooling timescale is relatively constant. As a result, the radius decreases more slowly as mass is lost, meaning that slightly more mass will be lost over a planet's history. This means that $f_{\mathrm{lost}}$ increases more than linearly with $\epsilon$. In turn this means that for fixed $f_{\mathrm{lost}}$, $F_{\mathrm{th}}$ decreases less than linearly with $\epsilon$. We can summarize both these trends in a single equation:

\begin{equation}\label{thresholdeq}
 F_{\mathrm{th}} = 0.5 F_{\mathrm{\oplus}} \left(\frac{M_{\mathrm{core}}}{M_{\mathrm{\oplus}}}\right)^{2 .4\pm0.4} \left(\frac{\epsilon}{0.1}\right)^{-0.7\pm0.1} 
\end{equation}

Equation (\ref{thresholdeq}) describes the results of our power-law fits for the location of $F_{\mathrm{th}}$. This is the incident flux a planet needs to receive from its parent star to remove half a planet's initial H/He envelope over the planet's lifetime, as a function of core mass $M_{\mathrm{core}}$, photo-evaporation efficiency $\epsilon$, and the flux that the Earth receives from the Sun $F_{\mathrm{\oplus}}$. This was derived by fitting to all the models in our parameter study and the error bars correspond to the 1$\sigma$ errors in those fits. Unfortunately, the dependence on initial composition is significantly more complicated and cannot be described by a simple power-law. 

In Figure \ref{fit1fig}c we plot $F_{\mathrm{th}}$ against the initial mass fraction in the H/He envelope at 10 Myr. For modest initial H/He envelopes, up to $\sim 60\%$ of the planet's total mass, increasing the H/He mass does not significantly affect $F_{\mathrm{th}}$. However, above this point the envelope's self gravity becomes sufficiently strong that density increases rapidly and it becomes increasingly difficult to remove any mass. At the same time, since these planets have such massive envelopes, removing a few Earth masses of H/He has a much smaller impact on their overall composition. Figure \ref{fit2fig} shows the other main effect of varying the initial composition. Here we show how the $M_{\mathrm{core}}-F_{\mathrm{th}}$ power-law index from Figure \ref{fit1fig}a depends on the initial composition. In general, as we increase the initial envelope mass, the radius-core mass relation becomes slightly steeper leading to a steepening of the $M_{\mathrm{core}}-F_{\mathrm{th}}$ relation.

In Figure \ref{fit3fig} we complete the picture by showing how the amount of mass lost depends on all these aspects. We plot $f_{\mathrm{lost}}$, the fraction of the initial envelope that is stripped after 5 Gyr, vs. the ratio of the incident flux and the threshold flux described by equation (\ref{thresholdeq}). The relation is roughly linear as described by equation (\ref{losteq}), where if $f_{\mathrm{lost}} \ge 1$ then the planet is completely stripped.

\begin{equation}\label{losteq}
f_{\mathrm{lost}} = 0.5 \left(\frac{F_{\mathrm{p}}}{F_{\mathrm{th}}} \right)^{1.1\pm0.3}
\end{equation}

Finally, in Figure \ref{fit4fig} we examine the quality of these simple analytic approximations compared to full results of our actual models. Here we plot the ratio of our full model and the analytic results of equations (\ref{thresholdeq}) and (\ref{losteq}) vs. the initial composition. The error bars represent the 1$\sigma$ scatter due to variations in $M_{\mathrm{core}}$, $\epsilon$, and $f_{\mathrm{lost}}$. For initial compositions that are less then $\sim 60\%$ H/He, the two generally agree with each other to within a factor of 2. For more massive initial envelopes, the analytic description breaks down and overstates a planet's vulnerability to mass loss. Also, it is important to keep in mind that these fitting equations are only rough approximations of the fully coupled evolution models and should not be used to make detailed predictions for individual planets. Nonetheless, these equations are valuable in understanding the qualitative behavior of our complete model and in making statistical comparisons to large populations of planets.

\subsection{Comparison to Observed Population}

Figures \ref{studyfig} and \ref{fit1fig} make a clear prediction about which planets should be most vulnerable to mass loss. In addition to the incident XUV flux that a planet receives, its mass loss history should depend strongly on the mass of its rock/iron core. Planets that are either highly irradiated or have low mass cores are more vulnerable to losing any primordial H/He envelope. Thus we expect that we should not find planets with H/He envelopes above the $F_{\mathrm{th}}-M_{\mathrm{core}}$ threshold relation in equation (\ref{thresholdeq}). Any planets that are well above this relation should either be rocky, or water worlds which are less vulnerable to mass loss \cp{Lopez2012}, or have H/He envelopes so large that equations (\ref{thresholdeq}) and (\ref{losteq}) breakdown.

Figure \ref{corefig} shows incident flux and core mass for all observed transiting planets with well defined masses less than 100 $M_{\mathrm{\oplus}}$ and radii less than 1.1 $R_{\mathrm{J}}$ from exoplanets.org \cp{Wright2011}. Two planets were removed by the radius cut since they are known to be inflated and it is impossible to accurately determine a heavy element mass for these planets. To calculate core masses for these 29 planets we ran water-free thermal evolution models without mass loss for each planet. The error bars on $M_{\mathrm{core}}$ include the observed uncertainties on mass, radius, and age as well as theoretical uncertainties on the iron fraction and thermal properties of the rocky core \cp{Lopez2012}. Each planet is color-coded by its current H/He envelope fraction. The red-brown open circles are planets that are consistent with being rocky today. The dashed black line shows the $F_{\mathrm{th}}-M_{\mathrm{core}}$ relation from equation (\ref{thresholdeq}) scaled up by a factor of two so that equation (\ref{losteq}) predicts complete stripping rather than only removing half the initial H/He envelope. Although the uncertainties are large, all the planets with substantial H/He envelopes are consistent with being to the right of this threshold. Of the six planets that lie to the left of the threshold, three (Kepler-10b, CoRoT-7b, and Kepler-20b) are consistent with being rocky and the other three with being water worlds (55 Cancri e, Kepler-18b, Kepler-20c).

\section{Effects on Planet Radii}
\subsection{Trends in the Radius-Flux Distribution?}

\begin{figure*}
  \begin{center}
    \includegraphics[width=7.in,height=4.81in]{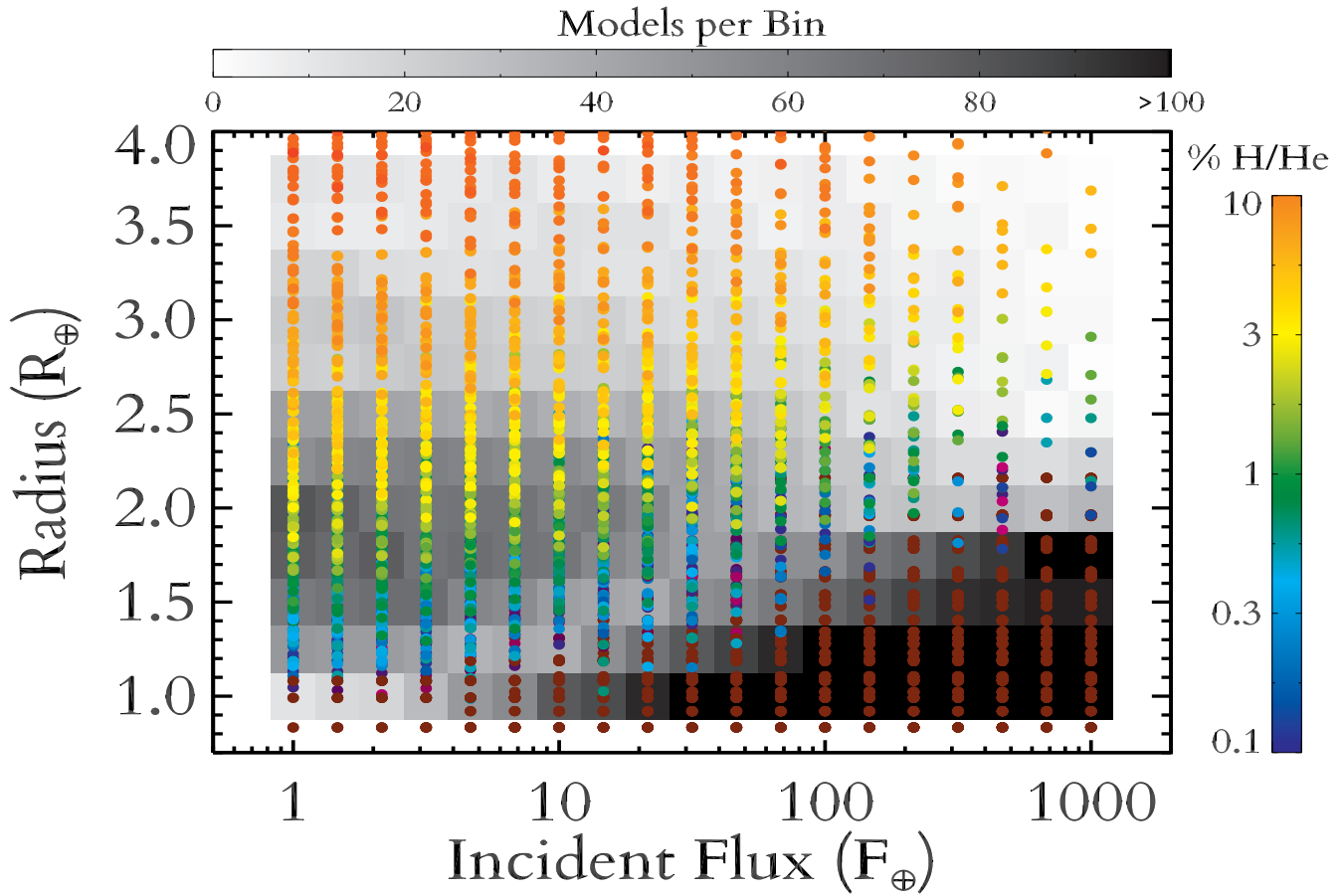}
  \end{center}
  \caption{Final planet radius vs. incident flux in our parameter study. The greyscale boxes indicate the frequency of models that end in each box due to thermal and mass loss evolution; i.e., black boxes contain many models and white boxes few. At high incident flux there is a strong decrease in the frequency of $\sim 1.8-4.0$ $R_{\mathrm{\oplus}}$ sub-Neptune sized planets and an increase in the frequency of $\lesssim 1.8$ $R_{\mathrm{\oplus}}$ rocky super-Earths. Points show the individual models, colored by their final H/He envelope fractions as result of photo-evaporation. The leftmost column of points at 1 $F_{\mathrm{\oplus}}$ closely approximates the distribution without any mass loss. Rust colored points in the bottom right correspond to rocky planets that have lost their envelopes. Just above these stripped cores there is a clear decrease in the frequency of models. Planets that enter this region have envelopes that are so small they tend to be stripped off completely. However, unlike \ct{Owen2013} we do not see a clear gap in planet radius. \label{gapfig}}
\end{figure*}

Thus far we have examined the effects of thermal and mass loss evolution on planet mass and composition. However, for the vast majority on {\it Kepler} candidates neither of these can be determined. As a result it is worth examining the predictions that our parameter study makes for planet radii. Figure \ref{gapfig} shows the radius-flux distribution of the models in our parameter study after 10 Gyr of thermal and mass loss evolution. As a reminder, we ran models on a log uniform grid of incident flux, core mass, and initial envelope fraction. Points show individual models colored by their final H/He fractions while the grayscale boxes show the density of points in the distribution. In order to directly compare with other recent studies described below, we limited ourselves to core masses $<$15 $M_{\mathrm{\oplus}}$, which also corresponds to the high end of core masses that will trigger runaway accretion \ct{Movshovitz2010}. Modest variations of the maximum core mass, from 10-20 $M_{\mathrm{\oplus}}$, do not qualitatively change any of the results described below. Planets with more massive cores, will have likely undergone runaway accretion and have final radii $>4$ $R_{\mathrm{\oplus}}$. In addition to varying the core mass, initial envelope fraction, and incident flux, we have also varied the iron fraction of the rocky core since this will smear out any trends with radius. We varied the iron fraction uniformly from pure silicate rock, to the maximum iron fraction allowed by collisional stripping, $\sim$60-80\% in this mass range \citep{Marcus2010}. For simplicity we restricted ourselves to only a single age and our standard mass loss efficiency 0.1.

Our parameter study here is not meant to produce a realistic radius-flux distribution. Our choice of a log uniform distribution in core mass, envelope fraction, and incident flux is meant to effectively probe the range of possible models. Nonetheless there are key features in Figure \ref{gapfig} that should be observable in the flux-radius distribution of {\it Kepler} candidates. First and foremost, there should be a decline in the rate of sub-Neptune sized planets, here defined as 1.8-4.0 $R_{\mathrm{\oplus}}$, at high incident fluxes due to photo-evaporation. In particular, these planets should become comparatively rare for $F_{\mathrm{p}}\gtrsim 100$ $F_{\mathrm{\oplus}}$, which corresponds to periods $\lesssim$10 days. Such an effect may have already been seen by \ct{Howard2011a} and \ct{Youdin2011}. At the same time there should be a corresponding increase in the frequency of rocky planets with $R_{\mathrm{p}}<$1.8 $R_{\mathrm{\oplus}}$, due to the large number of stripped cores.


\citet{Owen2013} recently performed a similar parameter study and predicted that there should be a significant gap in the radius distribution with few planets between 1.5 and 2.5 $R_{\mathrm{\oplus}}$. The origin of this gap is easy to understand. Planets in this size range will typically be $\sim$0.5\% H/He. Such small envelopes are highly vulnerable to photo-evaporation; even tiny mass loss rates $\sim$ $0.01$ $M_{\mathrm{\oplus}} \, Gyr^{-1}$ will be more than sufficient to strip these planets. Moreover, planets that formed with more substantial initial envelopes are unlikely to end up in this part of parameter space. This behavior was discussed at length in the context of evolution models for Kepler-11b in \ct{Lopez2012}. If a planet experiences enough mass loss to remove several percent of its mass, then it is much more likely to lose its envelope completely than to end up with an envelope that is $\sim$0.5\% H/He.

Nonetheless, we do not see a gap that is as clear cut as that found by \citet{Owen2013}. Instead we see a diagonal band in which models are relatively rare, although by no means excluded. This ``occurrence valley'' is typically 0.5 $R_{\mathrm{\oplus}}$ wide and occurs at slightly larger radii at higher incident fluxes. This is because in our parameter study we did not include planets that simply formed without any envelope at all. As a result, only planets that have lost their envelope to photo-evaporation end up being rocky. At the low flux end, only planets with initial envelopes $\sim$0.1\% H/He and the lowest core masses will lose their envelopes, resulting in relatively small stripped cores. On the other hand, at the high flux end even planets with initial envelopes up to $\sim$ 1\% H/He and core masses up to $\sim$10 $M_{\mathrm{\oplus}}$ are easily stripped. This removes somewhat larger H/He envelopes but results in larger stripped cores, moving the occurrence valley up to $\sim$2-2.5 $R_{\mathrm{\oplus}}$. 

However even in the middle of the diagonal occurrence valley, we find models from our study. Partially, this is due to varying the iron fraction of the cores, which smears out radius trends by $\sim$0.15 $R_{\mathrm{\oplus}}$. Mostly however, it is due to running a comprehensive parameter study that sampled a wide range of initial conditions. The \citet{Owen2013} study only tested five values for the core mass, without any initial compositions $<$1\% H/He. In contrast, Figure \ref{gapfig} includes 200 different combinations of core mass and initial composition. In the absence of photo-evaporation, this suite of models finely samples the entire range of radii from 1-4 $R_{\mathrm{\oplus}}$, as can be seen by the leftmost column in Figure \ref{gapfig}. We suggest that the reason \citet{Owen2013} see a clear gap in the radius-flux distribution is because their small sample of initial conditions do not adequately sample the parameter space.

Thus far no such gap has been has been seen in the observed distribution of planet radii. \ct{Fressin2013} and \ct{Petigura2013} recently performed careful studies of {\it Kepler} planet occurrence rates as a function of radius after correcting for false positives and the various selection effects. In both cases the studies find a flat occurrence rate below $\sim$2.8 $R_{\mathrm{\oplus}}$ with larger planets being significantly rarer. Both these studies span a wide range of periods, out to 85 days for \ct{Fressin2013} and 50 days for \ct{Petigura2013}, and use fairly wide radius bins. As a result, it is perhaps not surprising that they would not detect our relatively narrow occurrence valley. On the other hand, the flat occurrence rates found by \ct{Fressin2013} and \ct{Petigura2013} seem inconsistent with the wide gap proposed by \citet{Owen2013}. 

There are physical reasons why the occurrence valley seen in Figure \ref{gapfig} might not exist or be less pronounced. First, there could exist a large population of 1-10 $M_{\mathrm{\oplus}}$ rocky planets that simply formed without ever accreting a H/He envelope. This is possible if these planets formed through giant collisions after the disk had already dissipated \cp{Morbidelli2012}. This would introduce another population of 1-1.8 $R_{\mathrm{\oplus}}$ planets that would not show any strong dependence on incident flux due to photo-evaporation. Depending on how common these planets are, this could largely mask any gap in planet occurrence at radii $<$1.8 $R_{\mathrm{\oplus}}$, which corresponds to incident flux $\lesssim$100 $F_{\mathrm{\oplus}}$ or periods longer than $\sim$10 days. 

Second, if sub-Neptunes typically form beyond the snowline then in addition to rock, iron, and H/He, these planets could have large amounts of water and other volatile ices \cp{Rogers2011}. Much like varying the iron fraction of the core, varying the water fraction could wash out any trends in radius, but to a much greater extent. A 5 $M_{\mathrm{\oplus}}$ planet that is 50\% water will be $\sim$0.5 $R_{\mathrm{\oplus}}$ larger than one with an Earth-like composition \cp{Lopez2012}. Since our occurrence valley is only $\sim$0.5 $R_{\mathrm{\oplus}}$ wide, varying the water content sub-Neptunes from 0-50\% would completely eliminate any dip in planet occurance. {\it As a result, the presence or absence of such a dip is a useful test for whether sub-Neptunes form in situ without any water} \cp{Chiang2013,Hansen2012}, or migrate from beyond the snow-line with large amounts of water \cp{Rogers2011}. 

\subsection{Relative Sizes in Multi-Planet Systems}

Beyond simply explaining individual systems like Kepler-36, models of mass loss evolution may shed light on many of the puzzles of planet occurrence statistics. One such puzzle is that many of the {\it Kepler} multi-planet systems like Kepler-11, Kepler-18, \& Kepler-36 exhibit regular ordering of their radii; i.e. each planet tends to be larger than the one interior to it. This trend was recently quantified by \ct{Ciardi2012} which found that in pairs of planets from {\it Kepler} multi-planet systems there is a statistically significant tendency for the inner planet to be smaller than the outer planet. \ct{Ciardi2012} examined over 900 pairs of planets with periods ranging from 0.45 to 331 days and found that the inner planet is smaller in $\approx$60\% of planet pairs. Furthermore, they found that the fraction of planet pairs where the inner planet is smaller rises to $\sim$70\% when the both planets are within 20 days, and that the trend is only apparent for planets that are smaller than Neptune. Assuming that these planets also have masses less than Neptune, these are precisely the planets that should be vulnerable to photo-evaporation. We find that this trend can be explained as a natural result of photo-evaporation. Unless there is a strong tendency for planets on shorter periods to have more massive rocky cores, then inner planets should be substantially more vulnerable to photo-evaporative mass loss. 

On average we find that the inner planet in the \ct{Ciardi2012} sample receives 7.9 times more incident flux than the outer planet. Moreover, since this trend exists for pairs of planets orbiting the same parent star, we know that they receive the same XUV spectrum. Applying the scaling law derived in equation (\ref{losteq}), this implies that the inner planets in their sample should typically lose $\sim10$ times as much H/He. In \ct{Lopez2012} we showed that there is a near one to one correspondence between radius and the H/He mass fraction. As a result, this increased vulnerability to mass loss should naturally lead interior planets to have smaller radii. In addition, we would expect the fraction of pairs with smaller inner planets to rise at the shortest periods if these trends are in fact due to atmospheric mass loss. When both planet have periods shorter than $\sim$20 days, it is likely that both will be vulnerable to substantial mass loss and so there should be a larger impact on the relative radii.

However, there are several factors that can diminish the impact of mass loss on relative radii. Firstly, some planets will be rocky super-Earths without any volatile envelopes. Likewise our models predict that many planets on highly irradiated orbits should have envelopes completely stripped off. Of course, once a planet has lost its entire envelope its radius can not continue to shrink thus limiting any further differences in radii. Likewise, for planet pairs on less irradiated orbits, neither planet might be vulnerable to photo-evaporation. Also as we have already shown, large differences in rocky core mass can overwhelm differences in incident flux. All of the effects combine to limit the usefulness of radii alone to understand differences due to mass loss evolution and may explain why the trends seen by \ct{Ciardi2012} are relatively weak. Unfortunately, most of the planets in the Ciardi sample do not have mass measurements from radial velocity or TTVs, making it difficult to empirically test the importance of core mass on the trends they observe.

Currently, there are 16 pairs of {\it Kepler} planets where both planets have well determined masses and meet the SNR and impact parameter thresholds described in \ct{Ciardi2012}. These include planets in Kepler-9 \cp{Holman2010}, Kepler-10 \cp{Batalha2011}, Kepler-11 \cp{Lissauer2011a, Lissauer2013}, Kepler-18 \cp{Cochran2011}, Kepler-20 \cp{Fressin2011, Gautier2011}, and Kepler-36 \cp{Carter2012}. Of these 16 pairs, in four cases the inner planet is larger than the outer: Kepler-9b/c, Kepler-11c/f, Kepler-11d/f, Kepler-11e/f. In all four of these cases the inner is significantly more massive than the outer planet. With only four cases the trend is not yet statistically significant, however, the tendency for inner planets to be either smaller or significantly more massive is a robust prediction of our mass loss models. Whenever the inner planet in a multi-planet system has a larger radius, it should also be significantly more massive. There should not be any highly irradiated pairs of planets where the inner planet is less massive than the outer but where the inner planet has enough of its mass in a H/He envelope such that its radius larger.

It is worth mentioning that there are other processes in planet formation and evolution which could contribute to the trends seen by \ct{Ciardi2012}. \ct{Ikoma2012} showed that when super-Earths and sub-Neptunes form on highly irradiated orbits, the rate of H/He accretion is significantly slower when the local disk temperature is higher. Moreover, on short period orbits it is much easier for low mass planets to open a gap in the disk, which would also limit their envelope accretion \cp{Hansen2012}. Assuming that planets in the {\it Kepler} multis formed in the same ordering that they are in today, then inner planets should have had more difficulty in accreting large H/He envelopes. Likewise, given that proto-planetary disks evaporate from the inside out \cp{Calvet2000}, outer planets will have had more time to accrete an envelope. Nonetheless, given the critical mass loss timescale threshold that we identified observationally and theoretically in \ct{Lopez2012}, and a concurring view advanced by \ct{Wu2012}, it seems quite reasonable that photo-evaporative mass loss plays an important role in producing the trends among planet pairs seen by \ct{Ciardi2012}. Most likely, both planet formation and subsequent evolution combine to reduce the size of H/He envelopes for highly irradiated planets.

\section{Summary}
There is growing evidence from both models and observations that photo-evaporative mass loss plays an important role in the evolution of highly irradiated super-Earths and sub-Neptunes \cp{Baraffe2006,Jackson2012,Wu2012}. \ct{Lecavelier2007} first proposed that there could be a critical mass loss timescale in the observed population of hot Jupiters and Neptunes. In \ct{Lopez2012} we confirmed the existence this threshold for all planets with measured densities down to 2 $M_{\mathrm{\oplus}}$. Moreover, \ct{Lopez2012} showed that this critical mass loss timescale is naturally reproduced by our coupled thermal and mass loss evolution models. Likewise, this mass loss threshold is also reproduced by other models which fully solve the hydrodynamics of the mass loss wind \cp{Owen2012,Owen2013}. Here we have expanded upon the parameter study performed in \ct{Lopez2012} and shown in detail how mass loss history depends on incident flux, core mass, and mass loss efficiency. 

We have shown that in addition to the amount of XUV irradiation received by a planet, the mass of its rock/iron core plays a critical role in determining a planet's photo-evaporation history. Moreover we have shown that this provides a natural explanation for the large density contrast observed between Kepler-36 b\&c. In order to better understand the role of core mass, we performed an extensive parameter study and provided approximate scaling relations which can be used for estimates of whether mass loss has been important for detected planets. 

Further, we showed that the compositions of the observed population of transiting planets are consistent with our detailed models and these scaling relations. Finally, we showed that our coupled thermal and mass-loss evolution models make important predictions for the radius-flux distribution of {\it Kepler} candidates. In particular, we predict that sub-Neptune sized planets should become significantly less common at very short orbital periods. In addition, there may exist a narrow ``occurrence valley,'' which is a useful test for whether sub-Neptunes are formed in situ. We anticipate that future progress in this area will come from additional mass determinations of sub-Neptune size {\it Kepler} candidates, a better understanding of XUV fluxes from all types of stars as a function of age, and further progress in modeling mass loss efficiencies in the framework of 1D and 3D models.

\acknowledgements{We would like to thank Kevin Zahnle, Jack Lissauer, Yanqin Wu, and Erik Petigura for many helpful conversations. This research has made use of the Exoplanet Orbit Database and the Exoplanet Data Explorer at exoplanets.org." We acknowledge the support of NASA grant NNX09AC22G and NSF grant AST-1010017.}


\end{document}